\documentclass[aps,prl,twocolumn,nopacs,superscriptaddress]{revtex4}
\usepackage{graphicx}
\usepackage{verbatim}
\usepackage{mathtools}
\usepackage{amsmath}
\usepackage{sidecap}
\usepackage{epstopdf}
\usepackage{braket}
\usepackage{soul}
\usepackage[labelsep=period]{caption}

\begin{document}

\title{Robust surface states and coherence phenomena in magnetically alloyed SmB$_6$}

\author{Lin Miao}
\affiliation{School of Physics, Southeast University, Nanjing, 211189, China}
\author{Chul-Hee Min}
\affiliation{Experimentelle Physik VII and W\"{u}rzburg-Dresden Cluster of Excellence ct.qmat, Universit\"{a}t W\"{u}rzburg, Am Hubland, D-97074 W\"{u}rzburg, Germany}
\author{Yishuai Xu}
\affiliation{Department of Physics, New York University, New York, New York 10003, USA}
\author{Zengle Huang}
\affiliation{Rutgers Department of Physics and Astronomy, Rutgers University, Piscataway New Jersey 08854, USA}
\author{Erica C. Kotta}
\author{Rourav Basak}
\affiliation{Department of Physics, New York University, New York, New York 10003, USA}
\author{M. S. Song}
\author{B. Y. Kang}
\author{B. K. Cho}
\affiliation{School of Materials Science and Engineering, Gwangju Institute of Science and Technology (GIST), Gwangju 61005, Korea}
\author{K. Ki{\ss}ner}
\author{F. Reinert}
\affiliation{Experimentelle Physik VII and W\"{u}rzburg-Dresden Cluster of Excellence ct.qmat, Universit\"{a}t W\"{u}rzburg, Am Hubland, D-97074 W\"{u}rzburg, Germany}
\author{Turgut Yilmaz}
\author{Elio Vescovo}
\affiliation{National Synchrotron Light Source II, Brookhaven National Lab, Upton, New York 11973, USA}
\author{Yi-De Chuang}
\author{Weida Wu}
\affiliation{Rutgers Department of Physics and Astronomy, Rutgers University, Piscataway New Jersey 08854, USA}
\author{Jonathan D. Denlinger}
\affiliation{Advanced Light Source, Lawrence Berkeley National Laboratory, Berkeley, CA 94720, USA}
\author{L. Andrew Wray}
\email{lawray@nyu.edu}
\thanks{Corresponding author}
\affiliation{Department of Physics, New York University, New York, New York 10003, USA}

\begin{abstract}

Samarium hexaboride is a candidate for the topological Kondo insulator state, in which Kondo coherence is predicted to give rise to an insulating gap spanned by topological surface states. Here we investigate the surface and bulk electronic properties of magnetically alloyed Sm$_{1-x}$M$_x$B$_6$  (M=Ce, Eu), using angle-resolved photoemission spectroscopy (ARPES) and complementary characterization techniques. Remarkably, topologically nontrivial bulk and surface band structures are found to persist in highly modified samples with up to 30$\%$ Sm substitution, and with an antiferromagnetic ground state in the case of Eu doping. The results are interpreted in terms of a hierarchy of energy scales, in which surface state emergence is linked to the formation of a direct Kondo gap, while low temperature transport trends depend on the indirect gap.

\end{abstract}
% Include the date command, but leave its argument blank.

\date{\today}
\maketitle

The compound SmB$_6$ is a mixed-valence Kondo lattice system \cite{1,2,3} that has been under intensive study since 2010 as a strong candidate for the topological Kondo insulator (TKI) state \cite{4,5,6}. Angle-resolved photoemission (ARPES) measurements of the bulk band structure corroborate a TKI picture, in which dispersive 5$\it{d}$-orbital bands hybridize with semi-localized 4$\it{f}$-orbital states at the Fermi level to open a gap with topologically inverted symmetries \cite{7,8,9,10,11,12}. Topologically-associated surface states are seen to emerge as the gap opens at T$\sim$120K, and surface conductivity results in a low temperature (T$<$10K) resistivity plateau. The bulk insulating behavior and surface conductivity are strikingly robust against fractional changes in stoichiometry and non-magnetic alloying \cite{13,14,ImpurityPNAS}, however doping with magnetic elements (Ce, Gd) at a far lower $\sim$1$\%$ level eliminates signatures of surface conductivity \cite{13,14}, consistent with the expectation that magnetic disorder will introduce backscattering and Anderson localization to the topological surface state system \cite{TIreview1,TIreview2,19}. While these alloys represent fascinating extensions to the physics of the parent compound, Kondo latices are thought to be highly sensitive to disorder \cite{kondoHoles1,kondoHoles2,KondoHoleMIT}, and the fate of the TKI band structure in alloyed scenarios is unknown. Here, we report a comparative study of the electronic band structure, transport, and magnetic susceptibility properties of alloys incorporating small- and large-moment magnetic lanthanide elements as Sm$_{1-x}$M$_x$B$_6$ (M = Ce, Eu). A clean low-temperature gap in the bulk 5$\it{d}$ band is found to persist at the Fermi level up to the highest admixture levels tested (30$\%$ Ce, 20$\%$ Eu), and to be spanned by topologically-associated surface states. The emergence of Fermi level topological surface states in samples with an antiferromagnetic ground state (Sm$_{1-x}$Eu$_{x}$B$_6$, x$\geq$0.1) presents an antiferromagnetic topological insulator scenario \cite{15,MnBi2Te4_1} that has been much sought in recent years, with an advantage over other material candidates in that the topological band gap is positioned at the Fermi level. The physical conditions under which topological surface states emerge are found to remain tied to the \emph{direct} 5$\it{d}$ gap, and to allow for a broad range of low temperature resistivity trends defined by the \emph{indirect} Kondo gap \cite{16,17,18}.

%TO DO:
%***When discussing resistance to disorder, note that the system is not a conventional Kondo lattice, and the theoretical picture is far from complete. A generic model extending from SmB6 to CeB6 must be considered in the framework of \emph{multivalent} and \emph{dense} Kondo lattice scenarios, both of which present hard unsolved problems from the standpoint of theory and ARPES spectroscopy.

Single crystals of Sm$_{1-x}$M$_x$B$_6$ (M = Eu, Ce) were prepared by the alumina flux method, and details of the sample growth are described in the Supplemental Material (SM, Note 1 \cite{19}). Multiple characterization methods, including X-ray diffraction (XRD), scanning tunneling microscope (STM) and ultraviolet X-ray photoemission spectroscopy (UPS) were performed, revealing homogeneous alloying within single-phase crystals with stoichiometry-consistent substitution on the lanthanide site (SM, Note 2 \cite{19}). Most ARPES measurements were performed at beamline 4.0.3 at the Advanced Light Source, with a base pressure better than 5$\times$10$^{-11}$ Torr. The photon energy was set to h$\nu$=70 eV, corresponding to a bulk $\Gamma$-plane of the cubic Brillouin zone (see SM, Note 8 \cite{19} for other photon energies). Energy resolution was $\delta E \lesssim 10$ meV, and momentum resolution along the dispersive axis of measurement was $\delta k<3\times10^{-3}$ \AA$^{-1}$. Measurements of $k_z$ dependence for Sm$_{0.7}$Ce$_{0.3}$B$_6$ were performed at the NSLS-II ESM beamline, under approximately the same conditions. Samples were cleaved in situ at T$\sim$20K, using a top post glued to the (001) surface. All ARPES data were taken within 10 hours after cleavage, and band structure near the Fermi level was stable on this time scale (SM, Note 3 \cite{19}). Additional methods details for resistivity and magnetic susceptibility measurements are described in (SM, Note 4 \cite{19}). 

There are conflicting interpretations on whether a complete topological classification of SmB$_6$ surface states has been achieved \cite{20,21,22}, particularly with respect to weakly visible Fermi surfaces surrounding the Brillouin zone center. However, measurements consistently show a single surface state Fermi surface surrounding the surface $\overline{X}$ -point, with a contour that is not greatly influenced by surface termination \cite{7,8,9,10,11,12,20,21,22}. Spin-resolved investigations have found this state to be singly degenerate \cite{12,23,24}, creating a topologically nontrivial surface state scenario along the $\overline{X}$ -$\overline{M}$ axis (see Fig. 1(d) diagram, and SM, Note 5 \cite{19}).

\begin{figure}
\includegraphics[trim={0cm 0 0 0cm}, clip,width=8cm]{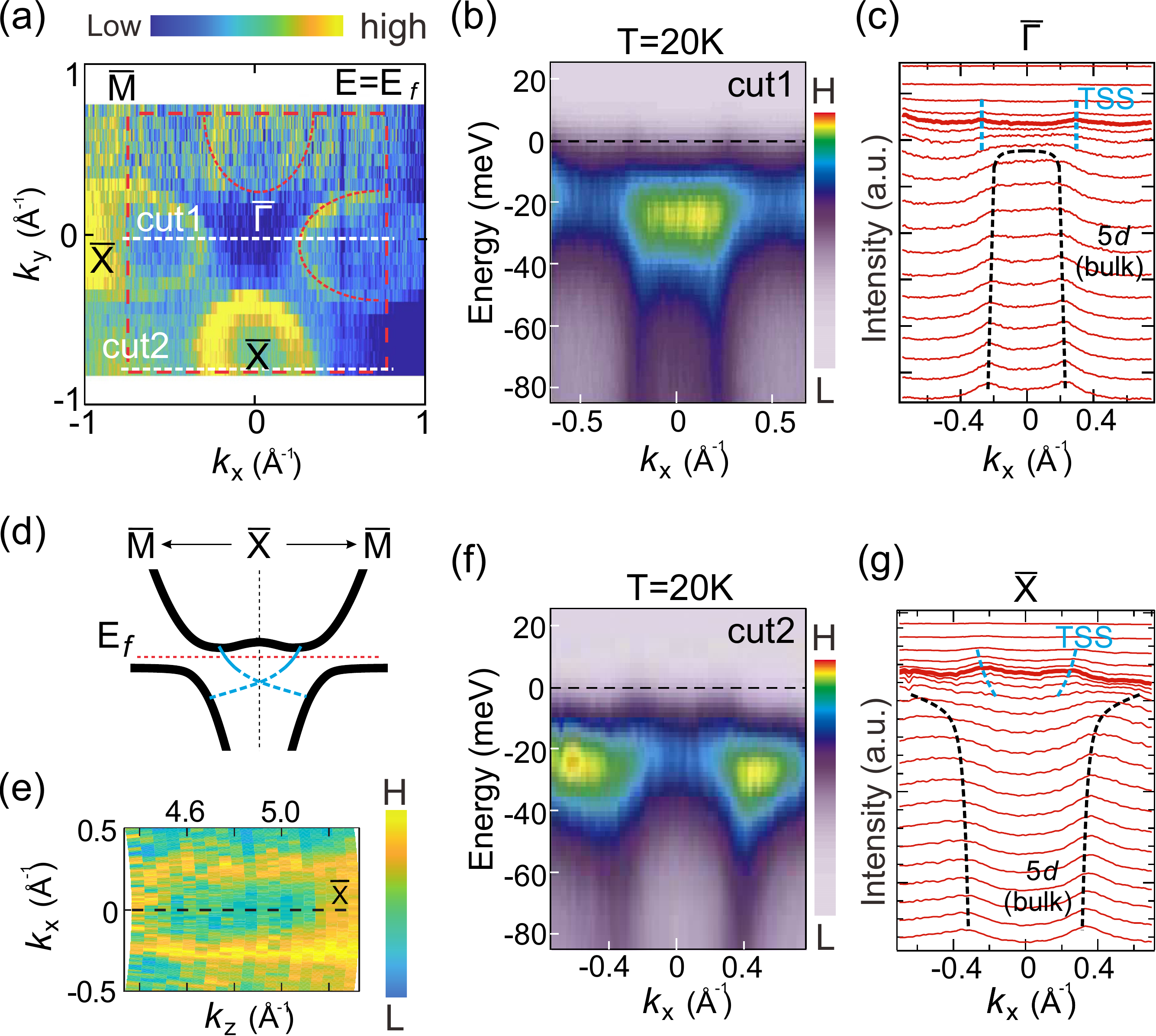} \captionsetup{font={small}}

\captionsetup{justification=raggedright,singlelinecheck=false}

\caption{\textbf{Surface states after 30$\%$ Ce doping.} (a) The Fermi surface of Sm$_{0.7}$Ce$_{0.3}$B$_6$, with dashed lines tracing ovoid surface state electron pockets. (b) A low-energy ARPES cut along the  $\overline{X}$ -$\overline{\Gamma}$ -$\overline{X}$ momentum axis (cut 1). (c) Surface (blue) and bulk (black) bands are traced on momentum distribution curves from panel (b), with an energy step of 6 meV. (d) A band structure diagram showing (blue) a topological surface state and (black) hybridization-gapped bulk bands along the $\overline{M}$ -$\overline{X}$ -$\overline{M}$ axis. The lower half of the surface Dirac cone is expected to be a weaker bulk-degenerate resonance state. (e) The two dimensionality of the surface state is visible in k$_z$-dependence of the cut 2 Fermi surface. (f-g) An ARPES cut along the $\overline{M}$ -$\overline{X}$ -$\overline{M}$ direction. The energy step in panel (g) is 5 meV. All measurements were performed at T=20K.}
\end{figure}

High-resolution ARPES experiments were performed on Sm$_{0.7}$Ce$_{0.3}$B$_6$ and Sm$_{0.8}$Eu$_{0.2}$B$_6$ to address the question of how this topological surface state responds to the altered chemical environments. For  Sm$_{0.7}$Ce$_{0.3}$B$_6$, low temperature (T=20K) measurements show a four-pocket Fermi surface with a long elliptical orientation along the $\overline{\Gamma}$ - $\overline{X}$ axis, where the Fermi momentum is marginally closer to $\overline{\Gamma}$ than to $\overline{X}$ (Fig. 1(a)). These states are two dimensional (Fig. 1(e); see also Supplementary Note 8 \cite{18}), and qualitatively identical to the $\overline{X}$ -point surface state Fermi surface observed at low temperature for undoped SmB$_6$. Fermi pockets surrounding the $\overline{\Gamma}$ -point are not seen, as is often the case for undoped SmB$_6$ under the same measurement conditions \cite{10,11}. Examining band dispersions along the $\overline{\Gamma}$ - $\overline{X}$ axis reveals steeply sloped bulk states (black dashed lines in Fig. 1(c)) from the $\overline{X}$-point 5$\it{d}$ pocket, which merge with a flat band associated with 4$\it{f}$ states. At the Fermi level, there is no feature at the extrapolated Fermi momentum of the 5$\it{d}$ bulk bands ($\sim$0.17{\AA}$^{-1}$ for Fig. 1(c), and $\sim$0.41{\AA}$^{-1}$ for Fig. 1(g)), indicating a well-defined bulk hybridization gap. The gap is crossed by highly dispersive surface states that are offset in momentum from the lower dispersion of the 5$\it{d}$ band (see guides to the eye in Fig. 1(c)) as is typically seen for undoped SmB$_6$. A similar scenario is seen along the $\overline{X}$ - $\overline{M}$ axis (Fig. 1(g)), with only a single surface state intersecting the Fermi level as required for the topological insulator state attributed to SmB$_6$.

\begin{figure*}
\includegraphics[trim={3cm 8.5cm 0 8.5cm}, clip, width=15cm]{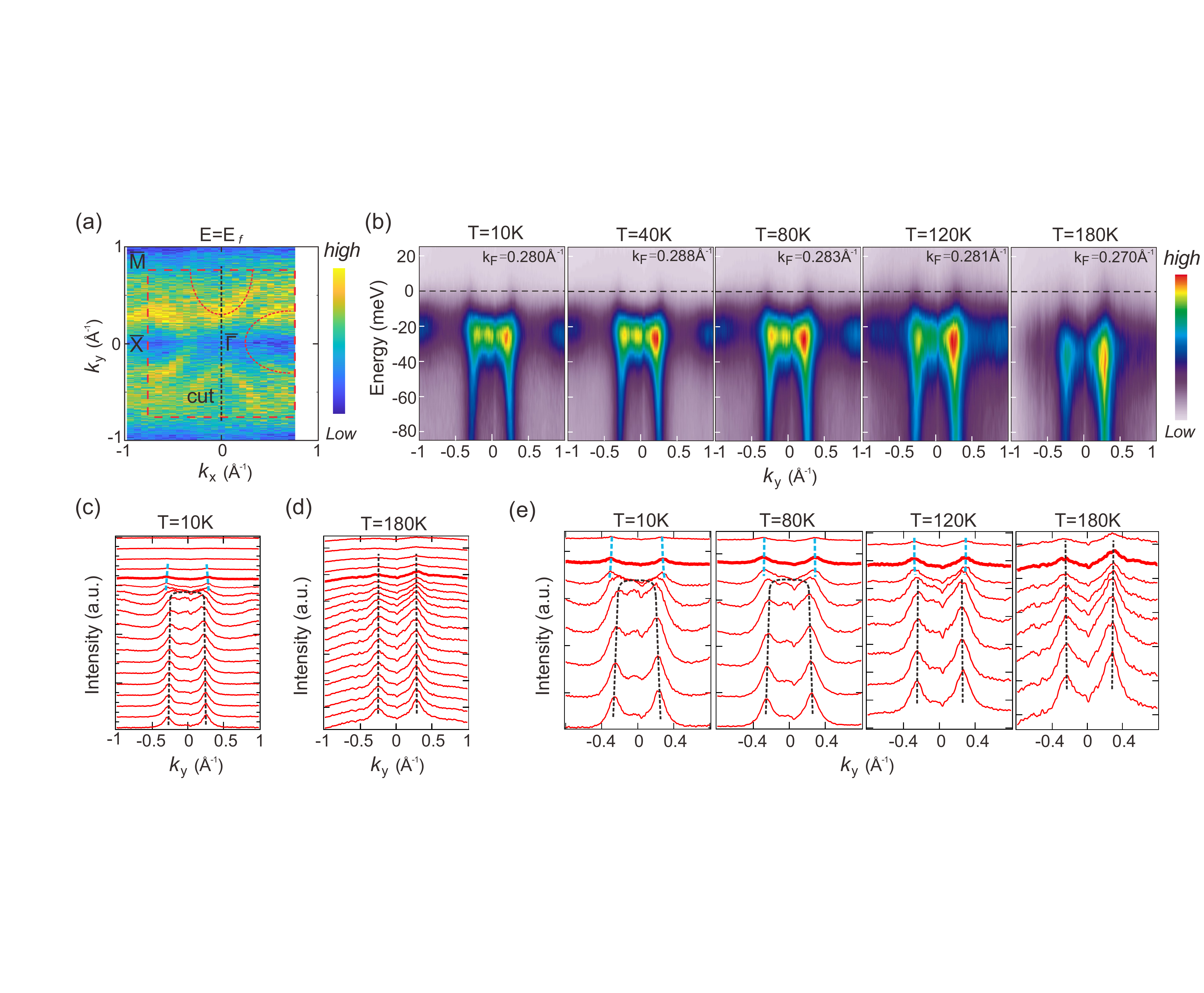} \captionsetup{font={small}}
\captionsetup{justification=raggedright,singlelinecheck=false}
\caption{\textbf{Surface state emergence after 20$\%$ Eu doping.} (a) The Fermi surface of Sm$_{0.8}$Eu$_{0.2}$B$_6$. (b) A low-energy ARPES cut taken along the high symmetry $\overline{X}$ -$\overline{\Gamma}$ -$\overline{X}$  direction (traced in (a)) is shown as a function of temperature, and annotated with the momentum-axis center of mass of Fermi level features. Raw data underlying the T=10K (c) and 180K (d) images are traced with surface (blue) and bulk (black) bands. The Fermi level is indicated with a thicker line, and the energy step is 6 meV. (e) Raw data curves within 50 meV of the Fermi level are shown for selected temperatures, with an energy step of 7 meV.}
\end{figure*}

The same features are seen in Sm$_{0.8}$Eu$_{0.2}$B$_6$ (Fig. 2(a-b)), however we find that the bulk electron pocket contours have shifted away from the Brillouin zone center by $\Delta$k = 0.07{\AA}$^{-1}$ relative to Sm$_{0.7}$Ce$_{0.3}$B$_6$. If taken to represent an isotropic fractional change in Fermi momentum, this indicates the addition of 0.2 holes per unit cell in the 5$\it{d}$ orbital. This difference in the bulk electronic structures can be understood by noting that the samarium sites in SmB$_6$ are mixed-valent with a roughly equal mixture of 4$\it{f}$$^5$ and 4$\it{f}$$^6$ configurations \cite{25}, whereas cerium and europium have strongly favored $\it{f}$-shell occupancies of 4$\it{f}$$^1$ (Ce$^{3+}$) and 4$\it{f}$$^7$ (Eu$^{2+}$) \cite{26,27} (see characterization in SM, Note 6  \cite{19}), and can be expected to contribute $\sim$50$\%$ fractional electron (Ce$^{3+}$) and hole (Eu$^{2+}$) doping, respectively, to the samarium sublattice. The observed difference in bulk 5$\it{d}$ dispersions accounts for roughly 80$\%$ of this nominal doping effect, suggesting that most of the doped charge resides in itinerant 5$\it{d}$ states, with just a few percent ($<$$\sim$5$\%$) of an electron or hole doped into the strongly correlated Sm 4$\it{f}$ sublattice.

%As in the Ce-doped and undoped cases, Sm$_{0.8}$Eu$_{0.2}$B$_6$ is seen to host four elliptical surface state electron pockets deriving from bands that disperse only above the 4$\it{f}$ band, and are offset from the bulk 5$\it{d}$ dispersion (Fig. 2(a-b)).

Raising the temperature of Sm$_{0.8}$Eu$_{0.2}$B$_6$ from 10K to 180K shows incremental broadening and a downward energetic shift of the flat 4$\it{f}$-state (Fig. 2(b)), as is also seen for undoped SmB$_6$ \cite{8,11}. The Fermi momentum of the in-gap feature declines at T$>$120K (see Fig. 2(b)), as the surface state (with larger Fermi momentum) vanishes and is replaced within the incoherent Kondo gap by a continuation of the bulk 5$\it{d}$ state (see Fig. 2(e) progression). The similarity of the low-temperature Fermi level band structure in these highly doped configurations is remarkable, but consistent with recent theory for mixed-valent systems \cite{41}, and with a majority of doped electrons occupying the itinerant 5$\it{d}$ states as attributed above.

\begin{figure}
\includegraphics[width=8cm]{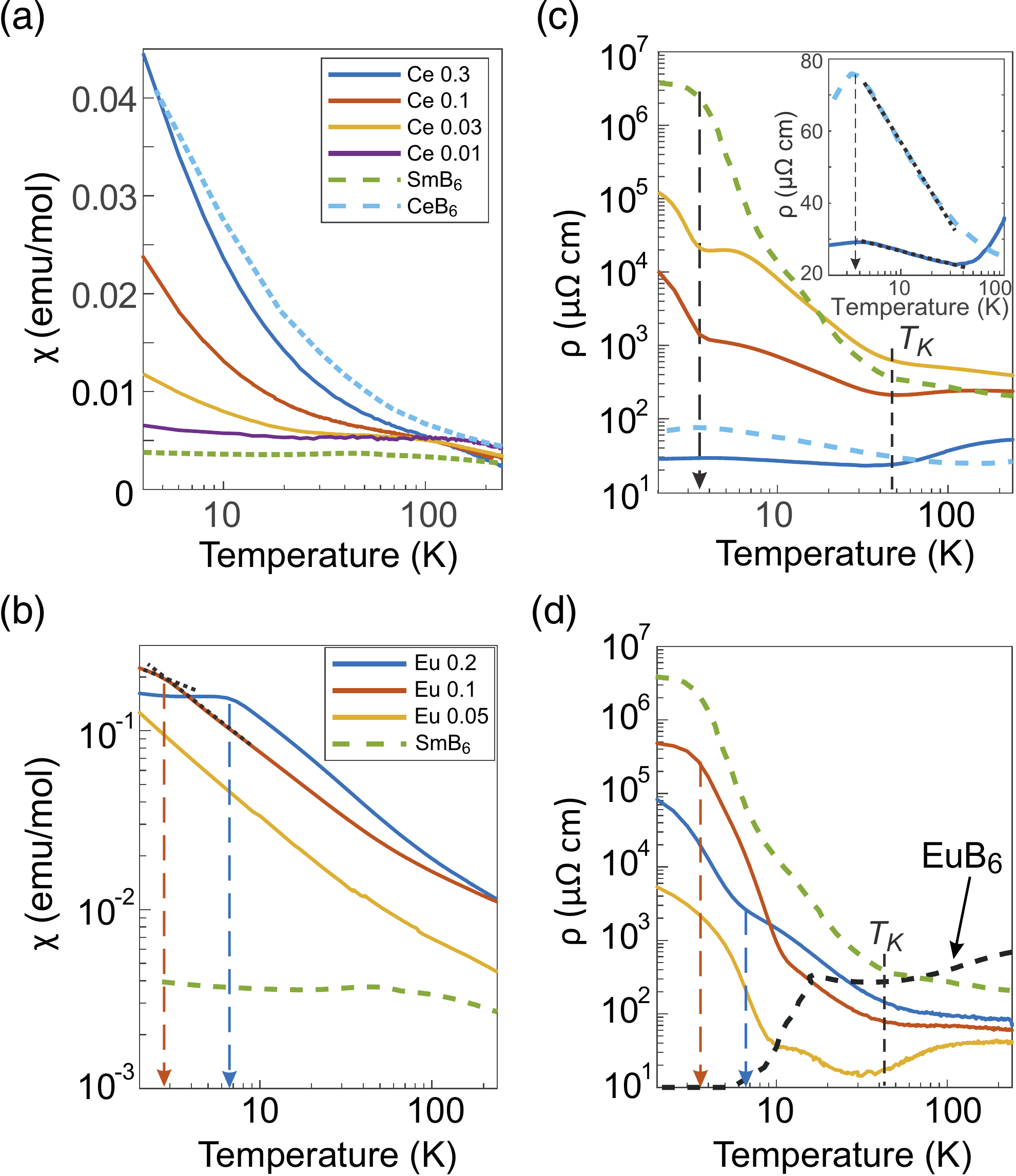} \captionsetup{font={small}}

\captionsetup{justification=raggedright,singlelinecheck=false}

\caption{\textbf{Resistivity and susceptibility of magnetically doped alloys.} (a)Magnetic susceptibility of Ce-alloyed SmB$_6$. Data for pristine CeB$_6$ and SmB$_6$ are extracted from refs. \cite{29} and \cite{45}. (b) Magnetic susceptibility of Eu-alloyed SmB$_6$. Drop-arrows mark the magnetic transition at 10$\%$ and 20$\%$ Eu doping levels. (c) Resistivity of Ce-alloyed SmB$_6$. A black drop-arrow marks the onset of the T$<$10K surface conductivity plateau for SmB$_6$, which happens to coincide with bulk-derived resistivity kinks in the alloyed samples. Data for pristine CeB$_6$ and SmB$_6$ are extracted from ref. \cite{46} and ref. \cite{47}, and an inset compares the resistivity trends for CeB$_6$ and Sm$_{0.7}$Ce$_{0.3}$B$_6$. (d) Resistivity of Eu-alloyed SmB$_6$, with data for pristine EuB$_6$ extracted from ref. \cite{30}. Red and blue arrows indicate magnetic transitions, as in panel (b).}
\end{figure}

Unlike these topological features, local moment physics and many-body ordering energetics cannot be unaffected by magnetic doping. As Ce doping level increases, the magnetic susceptibility ($\chi$-T) curves evolve rapidly and nonlinearly to strongly resemble the CeB$_6$ curve after just 30$\%$ substitution (Fig. 3(a)). Fitting inverse susceptibility (1/$\chi$-T) with a Curie-Weiss function in the high-temperature paramagnetic regime (SM, Note 7 \cite{19}) reveals effective local moments of 5.4$\mu$$_B$ (Sm$_{0.99}$Ce$_{0.01}$B$_6$), 5.15$\mu$$_B$ (Sm$_{0.97}$Ce$_{0.03}$B$_6$), 4.2$\mu$$_B$ (Sm$_{0.9}$Ce$_{0.1}$B$_6$), 3.15$\mu$$_B$ (Sm$_{0.7}$Ce$_{0.3}$B$_6$). The terminal value at 30$\%$ doping is remarkably close to the local moment of CeB$_6$, or the free-ion 4$\it{f}$$^1$ scenario, both of which are around 2.54$\mu$$_B$ \cite{28,29}.

Large-moment doping with Eu yields a similarly nonlinear trend, but in the opposite direction. Europium gravitates strongly to the half-filled large spin 4$\it{f}$$^7$ configuration favored by intra-atomic exchange interactions, with a large effective moment of ~8$\mu$$_B$ in EuB$_6$ \cite{30}. Partial substitution of samarium as Sm$_{1-x}$M$_x$B$_6$ induces antiferromagnetic order \cite{31}, which can be seen from the susceptibility kinks at T=2.8K (Sm$_{0.9}$Eu$_{0.1}$B$_6$) and T=7K (Sm$_{0.8}$Eu$_{0.2}$B$_6$) in Fig. 3(b). The effective moments extracted from inverse susceptibility for Eu-alloyed SmB$_6$ are 4.2$\mu$$_B$ (Sm$_{0.95}$Eu$_{0.05}$B$_6$), 7.1$\mu$$_B$ (Sm$_{0.9}$Eu$_{0.1}$B$_6$) and 6.1$\mu$$_B$ (Sm$_{0.8}$Eu$_{0.2}$B$_6$). The rapid change in local moments may be aided by a complementary effect from charge doping. Electron doping from Ce is expected to reduce local moment by promoting occupancy of the zero-moment ($^7${\bf F}$_0$) multiplet ground state of 4$\it{f}$$^6$ Sm, and hole doping from Eu will have the opposite effect by biasing the system towards large-moment 4$\it{f}$$^5$ configurations.

Transport trends are also very different between the Ce- and Eu-alloyed samples. The resistivity of undoped-SmB$_6$ is strikingly enhanced beneath the Kondo onset of T$\sim$50K (Fig. 3(c)), before flattening into a plateau associated with surface state conductivity \cite{13,32,33} at T$<$10K. Substituting 3$\%$ Sm with Ce causes T$\lesssim$5K bulk resistivity to be reduced by one to two orders of magnitude (Fig. 3(c)), an effect that seems counterintuitive in the context of adding defects, but matches expectations for an insulating Kondo lattice, as impurities may create in-gap states and suppress the coherence of the insulating Kondo band structure \cite{KondoHoleMIT,39}. The Kondo-regime resistivity of Sm$_{1-x}$Ce$_x$B$_6$ is suppressed further at higher doping levels, but retains an upturn beneath T$\lesssim$50K. For Sm$_{0.7}$Ce$_{0.3}$B$_6$, the trend beneath T$\lesssim$4K appears to be metallic (positively sloped), and resembles pure CeB$_6$ (see Fig. 3(c) inset).

The insulating character of Sm$_{1-x}$Eu$_x$B$_6$ (Fig. 3(d)) is far more robust, with a non-monotonic trend under doping that has been noted in previous literature \cite{31}. The alloys retain a characteristic Kondo insulating exponential trend (see SM, Fig. S9 \cite{19}). Mild inflections of resistivity are seen at the N\'eel transitions, but are not very distinguishable from other nonlinear details within the plot, and differ starkly from the dramatic change caused by ferromagnetism at T$_C$$\sim$12K in EuB$_6$ \cite{30}.

With respect to the trend toward metallicity in Ce- alloys, it is intriguing to note that pure CeB$_6$ also presents a 4$f$/5$d$ Kondo lattice scenario, with a superficially similar onset of Kondo-associated resistivity to SmB$_6$ at T$\lesssim$50K (Fig. 3(c)). However, the CeB$_6$ bulk band structure that emerges from low-temperature Kondo coherence is metallic due to the high density of Kondo-active large-moment sites \cite{34}, and the modest increase of resistivity as Kondo coherence sets in appears to represent the transition to a low-mobility heavy Fermion band structure \cite{35,36}. However, there is no fully established approach to modeling the band structure of a dense Kondo lattice like CeB$_6$, or a mixed-valent one like SmB$_6$. Theory for SmB$_6$ often builds from a computationally tractable single-particle band hybridization picture, however the monovalent nature ascribed to cerium in CeB$_6$ is much less compatible with such a picture, and the coincidence of Kondo temperature scales within resistivity measurements is suggestive that a complete theory for Sm$_{0.7}$Ce$_{0.3}$B$_6$ may need to go beyond a single-particle modeling basis.

From an empirical perspective, Sm$_{0.7}$Ce$_{0.3}$B$_6$ has superior T$<$50K conductivity to pure CeB$_6$ and preserves the topological surface state of SmB$_6$, and so appears to be a poor fit for the standard picture in which the metallicity of a disordered Kondo insulator comes from an in-gap impurity band \cite{KondoHoleMIT}. Examining the low temperature 4$f$ states in greater detail, we find that they are broader along the energy axis than those seen in SmB$_6$, with a peak width at half maximum (half-width) that is greater by $\sim$15-20 meV (Fig. 4(a)). This change is much larger than the $\sim$4 meV insulating gap attributed from the activation of resistivity in SmB$_6$ \cite{ImpurityPNAS}, and suggests that impurities may close the gap by introducing a broader continuum of diffusively conducting states. Similar 4$f$-associated states are seen by ARPES at low temperature in CeB$_6$, but are much broader \cite{37,38}, making it difficult to perform a fine comparison.

As temperature increases, a rapid thermal activation of 4$f$ half-width is seen for undoped SmB$_6$ and Sm$_{0.8}$Eu$_{0.2}$B$_6$ at T$\gtrsim$80K, and is consistent with the square of temperature (T$^2$) trend expected for Fermi liquid self-energy. In all cases, surface states and 4$f$/5$d$ hybridization are visible at 4$f$ half-width values less than 40 meV, found at temperatures T$\lesssim$120K. This phenomenology is consistent with recent theoretical investigations, which have proposed that the topological insulating gap of band insulator-like TKI systems may be highly robust against defects at the $\sim$20$\%$ level \cite{TKIholes,41}. This picture may also relate to the lack of surface states in ARPES studies of pure CeB$_6$ \cite{37,38}, which has a 4$\it{f}$-state width considerably larger than 40 meV.

The picture that emerges is one in which the topological surface states of SmB$_6$ alloys begin to emerge in concert with the direct 4$f$/5$d$ hybridization gap as coherence improves at T$\lesssim$120K. This matches expectations that the direct Kondo gap will define topology within a band picture (see Fig. 4(c)). The transition is unlikely to be sudden, as earlier studies on undoped SmB$_6$ have shown that the bulk bands evolve continuously throughout the onset of f-electron coherence \cite{11}. Surface states satisfy a nontrivial topology along the X-M axis, and are found near the outer boundary of the k$_z$-projected 5$d$ state continuum (Fig. 4(b)). The robustness of both surface states and the gap are remarkable given that SmB$_6$ may otherwise be unique as a TKI, and may be due to the unusual tendency of Sm towards \emph{mixed-valent} electronic configurations \cite{TKIholes,41} (see further discussion at the end of SM, Note 2 \cite{19}). Moreover, the crossover to a metallic (positively sloped) low temperature resistivity trend appears to relate to changes in the band structure on a smaller energy scale comparable to the $\sim$4 meV indirect band gap.

Taken collectively, these results show that TKI surface states can still emerge in alloys that deviate greatly from the parent compound SmB$_6$ in terms of metallicity, magnetic local moment, and magnetic ground state. This will enable the exploration of TKI surface physics in new physical regimes, and supports recent theoretical modeling, which has suggested that topological band gap and surface state properties of SmB$_6$-like TKI materials will be far more robust against disorder than is typically expected for Kondo lattice systems \cite{TKIholes,41}. Comparing these alloys with undoped SmB$_6$ and CeB$_6$, the emergence of TKI surface states is found to consistently coincide with the  opening of a direct band gap in the bulk 5$d$ spectral function, which in turn is associated with 4$f$-state half-widths falling beneath a critical $\lesssim$40 meV threshold. Local moments are seen to vary by a factor of two, from roughly 3.15$\mu$$_B$ in Sm$_{0.7}$Ce$_{0.3}$B$_6$ to over 6$\mu$$_B$ in Sm$_{1-x}$Eu$_x$B$_6$ (x=0.1, 0.2), and the differing 4$\it{f}$-electron energetics appear to result in contrasting trends towards metallic-like and insulating-like Kondo lattice scenarios for Ce- and Eu-alloys, respectively. The insulating-like scenario realized by Sm$_{0.8}$Eu$_{0.2}$B$_6$ places topological surface states at the Fermi level in a system with T$_N$=7K antiferromagnetism, presenting a remarkably clean candidate system for the much-sought antiferromagnetic TI state.

%The formation of coherent Kondo singlets is more challenging to achieve as local moment density increases in Ce-rich compositions, and the Kondo coherent low-temperature band structure (should it appear) is expected to include larger dispersions that make the indirect Kondo gap negative and thus render the system metallic \cite{18} (see the 3$^{rd}$ panel of Fig. 4(c)).
%***When discussing resistance to disorder, note that the system is not a conventional Kondo lattice, and the theoretical picture is far from complete. A generic model extending from SmB6 to CeB6 must be considered in the framework of \emph{multivalent} and \emph{dense} Kondo lattice scenarios, both of which present hard unsolved problems from the standpoint of theory and ARPES spectroscopy.

%-complicated modeling regime
%-temperature/energy scales not distinguished for Kondo lattice physics; renormalized band physics; tendency to model in band picture is largely because there is no fully established approach to modeling band structure for the dense Kondo lattice and mixed-valent Kondo lattice scenarios
%-within this, a big picture

% trim={0.5cm 0.3cm 0.8cm 0.5cm}
\begin{figure}
\includegraphics[width=8.7cm]{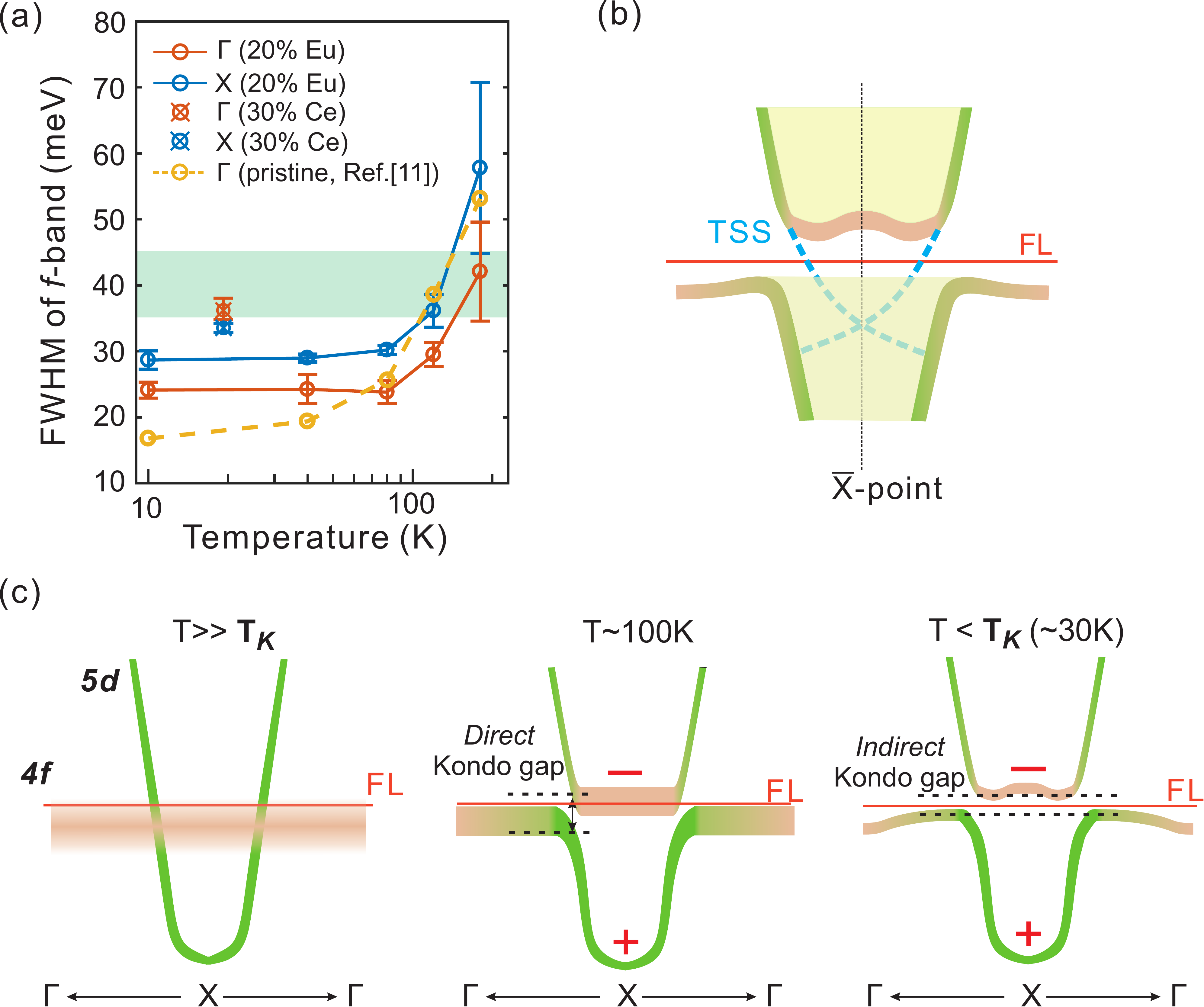} \captionsetup{font={small}}

\captionsetup{justification=raggedright,singlelinecheck=false}

\caption{\textbf{Coherence, hybridization and surface state emergence.} (a) The 4$\it{f}$ band feature width at half maximum in alloys is evaluated from a Voigt fit and compared with pristine SmB$_6$ as a function of temperature. A shaded region indicates the threshold below which surface states become visible to ARPES. (b) A schematic showing surface states spanning the hybridization gap. Bulk states with zero z-axis momentum are drawn as solid lines, and the k$_z$-projected bulk state continuum is shown with light yellow shading. (c) Diagrams show the temperature-resolved evolution of TKI bulk band structure intersecting the X-point. (middle) As coherence increases with the lowering of temperature, topologically inverted direct hybridization gaps open at band crossing points. (right) As temperature decreases further, the presence or absence of an indirect gap within the Kondo band structure defines the trend of resistivity. The topologically inverted parity symmetries are indicated by $'$+$'$ ($\it{d}$-orbital) and $'$-$'$ ($\it{f}$-orbital) signs}
\end{figure}

\section*{ACKNOWLEDGEMENTS}

%XAS, XPS and ARPES were performed 

Spectroscopic measurements in the main text were performed at the Advanced Light Source, supported by the Director, Office of Science, Office of Basic Energy Sciences, of the U.S. Department of Energy under Contract No. DE-AC02-05CH11231. Operation of the ESM beamline at the National Synchrotron Light Source is supported by DOE Office of Science User Facility Program operated for the DOE Office of Science by Brookhaven National Laboratory under Contract no. DE-AC02-98CH10886. Work at NYU was supported by the MRSEC Program of the National Science Foundation under Grant Number DMR-1420073. The STM work at Rutgers is supported by NSF under grant DMR-1506618. C.-H. Min, K. Ki{\ss}ner, and F. Reinert acknowledge financial support from the DFG through SFB1170 `tocotronics' and the W\"{u}rzburg-Dresden Cluster of Excellence on Complexity and Topology in Quantum Matter - ct.qmat (EXC 2147, project-id 39085490). M.S. Song, B.Y. Kang and B. K. Cho were supported by National Research Foundation of Korea (NRF), funded by the Ministry of Science, ICT \& Future Planning (No. NRF-2017R1A2B2008538). L. Miao is supported by the National Natural Science Foundation of China (Grants No. U2032156 and No. 12004071) and Natural Science Foundation of Jiangsu Province, China (Grant No. BK20200348).

\end{document}